\documentclass[runningheads]{llncs}

\usepackage[T1]{fontenc}
\usepackage{graphicx}
\usepackage{amsmath}
\usepackage[table]{xcolor}
\usepackage{multirow}

\begin{document}

\title{STAR-NT: Spatiotemporal Acceleration of Real-Time Neural
Transparency Rendering}

\author{Grigoris Tsopouridis\inst{1}\orcidID{0000-0001-8033-5481} \and
Christos Georgiou-Mousses\inst{1} \orcidID{0009-0004-4912-3332}\and
Aris Panagiotidis\inst{1}\orcidID{0009-0008-2964-4359} \and 
Andreas Vasilakis\inst{2}\orcidID{0000-0001-6895-3324}
David Corrigan\inst{3} \and Tobias A. Franke\inst{3}\orcidID{0009-0003-9025-9277} \and Aleksei Gorbonosov\inst{3} \and Andrei Astapov\inst{3}
\and 
Ioannis Fudos\inst{1}\orcidID{0000-0002-4137-0986}
}

\authorrunning{G. Tsopouridis et al.}
\titlerunning{STAR-NT: Spatiotemporal Acceleration for Neural OIT}
\institute{University of Ioannina, Greece
\email{\{g.tsopouridis,fudos\}@uoi.gr}
\and
Athens University of Economics and Business, Greece\\
\and
Huawei Ireland Research Center, Ireland
}
\maketitle              
\begin{abstract}
Neural order-independent transparency delivers high-quality rendering of overlapping transparent surfaces, but its geometry passes and network input generation remain costly, particularly on mobile and legacy hardware. We present a spatiotemporal acceleration framework that exploits spatial and temporal coherence to reduce this overhead while preserving visual quality. Spatially, we use adaptive quadtree-based screen-space subdivision to scale geometry pass resolution according to local color variance. Temporally, selected frames reuse the previous transparency result through depth-based reprojection instead of full rendering. Together, these optimizations reduce rendering cost and integrate efficiently into existing real-time rendering pipelines.

\end{abstract}

\section{Introduction}

Order-Independent Transparency (OIT) is a fundamental challenge in real-time computer graphics that enables correct rendering of transparent objects without depth-based sorting. Since correct alpha blending~\cite{porter84} requires sorting, it is impractical for complex intersecting geometry, and traditional object-sorting approaches may produce artifacts under multiple overlapping transparent surfaces. OIT avoids these limitations by compositing transparency independently of rendering order, improving visual realism in games, simulations, and visual effects.


Traditional OIT methods are typically divided into \textit{exact} and \textit{approximate} approaches. Exact methods, such as depth peeling~\cite{Everitt2001,Bavoil2008OrderIT} and the A-buffer~\cite{Carpenter1984,yang2010}, correctly capture all fragments per pixel but incur high and scene-dependent memory and performance costs, making them unsuitable for real-time applications. Approximate methods, including Weighted Blended OIT~\cite{McGuire2013} and Moment-Based transparency~\cite{Munstermann2018,Sharpe2018}, provide bounded cost and predictable performance, but introduce visual artifacts due to compositing approximations.


Recent work has explored neural OIT, where learned models approximate transparency compositing with bounded resources. DFAOIT~\cite{tsopouridis2024dfa} uses two depth-peeled layers and per-pixel statistics to predict final pixel colors via a lightweight neural network. While achieving high visual quality with constant memory usage, it incurs significant overhead from per-pixel inference and input generation, limiting performance on low-end and legacy hardware.
In this work, we present a dual-domain acceleration framework for neural OIT that reduces computational overhead through spatial and temporal optimizations. Our key insight is that geometry pass cost can be reduced by exploiting (i) spatial redundancy, where low-complexity may not require full-resolution evaluation, (ii) temporal coherence, where consecutive frames share structure enabling transparency reprojection, and (iii) a streamlined network design that predicts only tail transparency color for improved efficiency. This paper makes the following technical contributions:

\begin{itemize}

\item A quadtree-based screen-space adaptive subdivision scheme that uses local color variance to adjust geometry pass resolution, reducing shading and overdraw. We additionally employ depth-based temporal reprojection to reuse previous transparency results via frame-to-frame coherence.

\item A redesigned neural network with a reduced input feature set, addressing limitations of DFAOIT~\cite{tsopouridis2024dfa} while improving inference efficiency and maintaining comparable or better quality than non-neural OIT methods.

\item A unified framework combining spatial and temporal optimizations, achieving an average 39\% speedup while preserving order-independence and requiring no scene-specific training or preprocessing.

\end{itemize}

Our approach preserves order-independence and requires no preprocessing, scene-specific training, or modification to the underlying OIT method. Operating at the rendering-pipeline level, it can be integrated into existing neural transparency techniques, improving their suitability for interactive applications such as medical visualization, cultural heritage, and mobile rendering.
\section{Related Work}

\textbf{Multifragment Rendering.} Early OIT methods focused on exact solutions. The A-Buffer~\cite{Carpenter1984} stores and sorts all per-pixel fragments before alpha compositing~\cite{porter84}, achieving correct results at the cost of unbounded memory. Depth peeling~\cite{Bavoil2008OrderIT} extracts and blends layers iteratively, but its cost scales with the number of visible layers. Bounded variants such as the k-buffer~\cite{kbuffer} limit per-pixel storage for predictable performance, trading accuracy for efficiency. Despite these improvements, exact multifragment methods remain costly in scenes with high depth complexity.


\textbf{Blended Transparency.} Approximate OIT methods reduce the memory and computational costs of exact approaches by avoiding fragment sorting and multi-pass processing. Weighted sum~\cite{Meshkin2007} and weighted average~\cite{Bavoil2008OrderIT} accumulate fragment contributions with a simple color weighted sum or average, achieving high performance at the expense of visual artifacts. Weighted Blended OIT (WBOIT)~\cite{McGuire2013} improves quality using empirical, user-defined depth and opacity-based weights, but its heuristic formulation can produce artifacts in scenes with high opacity variation or depth complexity. Layered WBOIT~\cite{friederichs2021layered} further improves quality by applying blending per depth bin, at increased computational cost.

\textbf{Hybrid Transparency.} Hybrid transparency combines exact and approximate OIT to balance visual accuracy and computational efficiency. These methods typically compute a small set of foreground layers exactly while approximating the remaining fragments. Maule et al.~\cite{Maule2013} use a k-buffer for the foremost fragments and approximate the tail using weighted blending. This improves accuracy for visually important layers while avoiding the unbounded memory cost of A-buffer and the high overhead of depth peeling. However, performance and quality remain sensitive to the chosen number of exact layers.


\textbf{Neural Transparency.} Recently, a number of methods have explored the use of neural networks to enhance or replace the conventional transparency pipeline~\cite{tsopouridis2025tut}. Deep and Fast Approximate OIT~\cite{tsopouridis2024dfa} uses a lightweight neural network to predict the final composite per-pixel colors from their color and opacity statistics, allowing bounded-cost per-pixel inference suitable for real-time use. Neural Moment Transparency~\cite{Tsopouridis2024NeuralMT} follows a hybrid strategy that improves quality over moment-based transparency~\cite{Munstermann2018} by replacing selected stages with learned models, but presents a higher computational cost due to additional moment computation and reconstruction steps. Deep Hybrid Order-Independent Transparency~\cite{Tsopouridis2022} integrates a neural pixel-importance estimator into a variable k-buffer pipeline, achieving higher quality at the cost of significantly increased computational and memory demands. Neural networks have also been used to approximate screen-space effects such as ambient occlusion, depth of field, indirect illumination~\cite{Nalbach2017}, and shadow mapping~\cite{Datta2022}, achieving high-quality results at reduced computational cost.


\textbf{Accelerating Rendering.} 
Recent rendering advances use tile-based processing and coarse pixel shading~\cite{coarsepixelshading} to improve performance under bandwidth and power constraints. Tile-based methods partition the screen into spatial regions to improve cache efficiency and reduce memory traffic, lowering shading rates in the GPU pipeline based on the tile resolution. Temporal methods such as TAA~\cite{Karis2014} and neural frame interpolation~\cite{neuralinterp} exploit frame-to-frame coherence to reduce computation. We apply the same principles to accelerate neural OIT while preserving visual quality.

\section{
STAR-NT: Spatial-Temporal Acceleration for Real-time Neural
Transparency}
\begin{figure}[!h]
    \centering
    \includegraphics[width=0.70\linewidth]{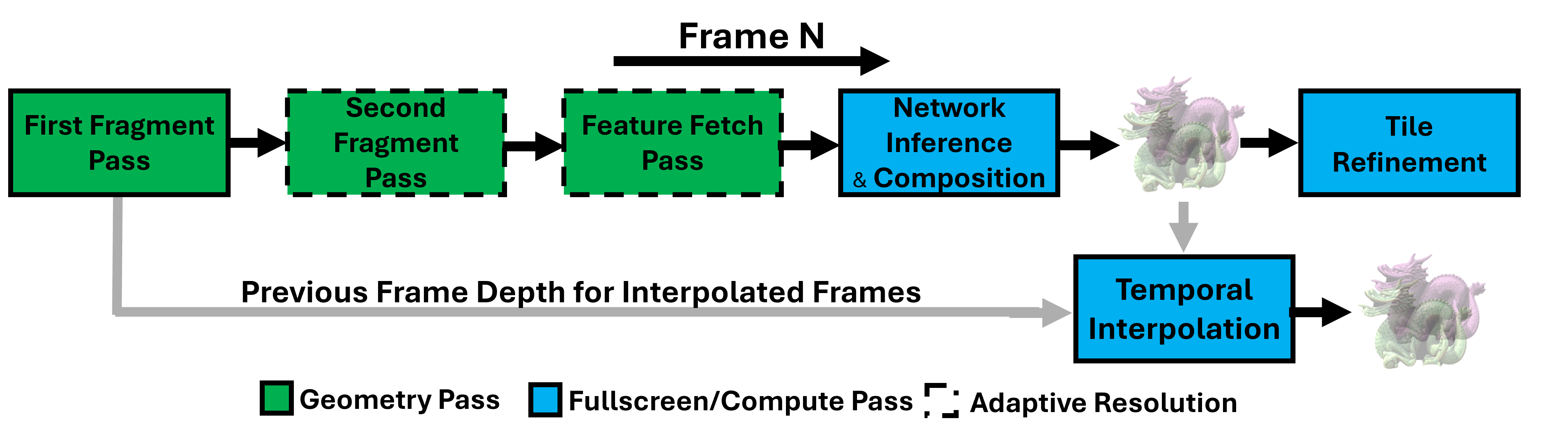}
    \caption{Geometry passes execute at adaptively reduced resolutions while inference and composition run at full resolution, driving tile refinement for the subsequent frame. First fragment depth  is reused for temporal reprojection (grey arrow). }
    \label{fig:passes}
\end{figure}
We present an acceleration framework for neural transparency rendering that combines adaptive resolution control with temporal frame interpolation. An adaptive screen-space tile subdivision scheme, guided by a transparency color difference metric, dynamically adjusts the resolution of geometry passes, reducing rendering cost while preserving visual fidelity. Tiles computed in frame $F_t$ guide the adaptive resolution of frame $F_{t+1}$. In addition, we utilize a reprojection-based temporal interpolation stage that exploits temporal coherence under smooth camera motion to amortize computation across frames. Finally, we refine the neural network input features to improve performance and address some of the limitations of DFAOIT~\cite{tsopouridis2024dfa} (See supplemental material). We build upon the DFAOIT low-end version, using three geometry passes (Fig. \ref{fig:passes}), avoiding interlocks, making this method suitable for mobile and legacy hardware. The first two passes compute the first and the second closest fragments, while the third pass computes the input features of our neural network. We utilize compute-shader-based tiled processing, mixed-resolution rendering, achieving high-quality transparency rendering with a GPU memory footprint comparable to DFAOIT (75 MB at 1080p). Shader source code is available online: https://github.com/gtsopus/STAR-NT.

\subsection{Adaptive Screen-space Tiles}
\label{adaptivetiles}

The rendering cost of neural transparency is dominated by geometry passes that rasterize the full scene multiple times. This cost is highly non-uniform across the screen as regions with dense, overlapping, or high-contrast transparent geometry demand greater fidelity than flat or sparsely covered areas. Rather than rendering all passes at full resolution, we employ an adaptive, bitmask-based screen-space tile subdivision system that works as a quality metric to selectively drive the resolution of each geometry pass, based on the corner-sample color variation measure $D_i$ (See supplemental material).

The proposed adaptive tiling scheme dynamically subdivides the screen into variable-resolution regions according to image-space color variation measured from the neural OIT inference output. Starting from \(16\times16\) base tiles, the method constructs a four-level quadtree hierarchy down to \(2\times2\) regions, where each tile encodes its refinement state compactly within a 21-bit mask stored in a single 32-bit integer. Refinement is evaluated in a post-transparency compute pass using only the four corner samples of each node, with local variation estimated through the mean pairwise Euclidean color distance $D_i$. This constant-cost \(O(1)\) evaluation enables efficient split and merge decisions driven by thresholds, with the split thresholds being smaller than the merge thresholds, reducing temporal instability while preserving structural consistency through top-down refinement. The resulting hierarchy is reused across frames to guide spatially adaptive rendering resolution, achieving efficient detail allocation while avoiding the bandwidth overhead of full-tile analysis. To reduce perceptual artifacts, neighboring tiles may differ by at most one refinement level, and tiles may change by only one level per frame. Since refinement levels are limited to \(\{0,1,2,3\}\), two relaxation iterations (\(k=2\)) are sufficient to enforce spatial consistency across the grid.



The resulting tile hierarchy is then used in the subsequent frame. In particular, the number of active tiles at each level serves as a demand indicator for our adaptive transparency rendering (Sec.~\ref{adaptiveresolution}), guiding the second and third pass resolutions.

\subsection{Adaptive Tile-Guided Resolution}
\label{adaptiveresolution}

The rendering cost of each geometry pass is proportional to the resolution at which it is executed. For transparency, we observe that the visual contribution of each successive pass diminishes under the Porter-Duff over-compositing operator~\cite{porter84}, as the color contribution of layer $p$ is attenuated by the accumulated transmittance of all preceding layers,

\begin{equation}
    C_{\mathrm{out}} = \sum_{p=1}^{P} C_p \cdot \prod_{q=1}^{p-1}(1 - \alpha_q),
\end{equation}

\noindent where $C_p$ and $\alpha_p$ are the color and opacity of layer $p$ respectively. As depth increases, the product $\prod_{q}(1-\alpha_q)$  decreases, attenuating the contribution of deeper layers to the final result. This motivates rendering successive passes at progressively reduced resolution, as the loss in spatial fidelity is masked by the reduced transmittance of the affected layers.


The first pass is always rendered at full resolution since it captures the
foremost transparent layer, which contributes most strongly to the final
image and contains the most perceptually important detail. Reducing its
resolution introduces visible artifacts that later passes cannot recover.

Passes 2 and 3 are rendered at resolutions determined by the scene complexity
index \(\bar{w}\), computed from the tile
level histogram of the previous frame (Section~\ref{adaptivetiles}). The
index is defined as the level-weighted mean of the occupied tile distribution:
$\bar{w} = \frac{\sum_{\ell=0}^{3} \ell \cdot N_\ell}{\sum_{\ell=0}^{3} N_\ell}$, where $N_\ell$ denotes the number of occupied tiles at level $\ell$. The index $\bar{w} \in [0, 3]$ varies continuously as the tile distribution shifts along scene and camera movements. Each pass maps $\bar{w}$ to a render resolution scale through a pass-specific exponential function:

\begin{equation}
    s_{p} = \mathrm{clamp}\!\left(2^{\bar{w} - \delta_p},\; 0.25,\; 1.0\right), \\
    \delta_2 = 2.2, \, \delta_3 = 3.0
    \label{eq:scale}
\end{equation}


The base-2 exponential form is a natural choice given the power-of-two structure of the subdivision hierarchy: in the unclamped range, a unit increase in $\bar{w}$ corresponds to a doubling of the render scale. The offset $\delta_p$ sets the crossover point at which pass $p$ attains full resolution: $s_2 = 1$ when $\bar{w} = 2.2$, and $s_3 = 1$ when $\bar{w} =
3.0$.  Therefore, $\delta_p$ determines the trade-off between performance and quality.  Pass 3 is assigned a higher offset, rendering it at lower resolution than Pass 2 under equivalent complexity, as its tail color statistics are used  by the neural network, which can compensate for partially degraded inputs. Both scales are clamped to a minimum of $0.25$, below which quality degrades unacceptably with diminishing performance returns. Pass 3 resolution changes may cause minor temporal artifacts, but hysteresis in split/merge thresholds prevents oscillations, and large changes only occur during major scene or camera shifts.


Since neural network inference operates at full resolution, the outputs of Passes 2 and 3 are first upscaled to framebuffer resolution using bilinear interpolation. The upscaled Pass 3 buffers are then provided as neural network inputs (Sec.~\ref{neuralnetwork}) and composited together with the full-resolution Pass 1 output, the upscaled Pass 2 result, and the background
color.

\subsection{Neural Network Improvements}
\label{neuralnetwork}
\begin{figure}
    \centering
    \includegraphics[width=0.45\linewidth]{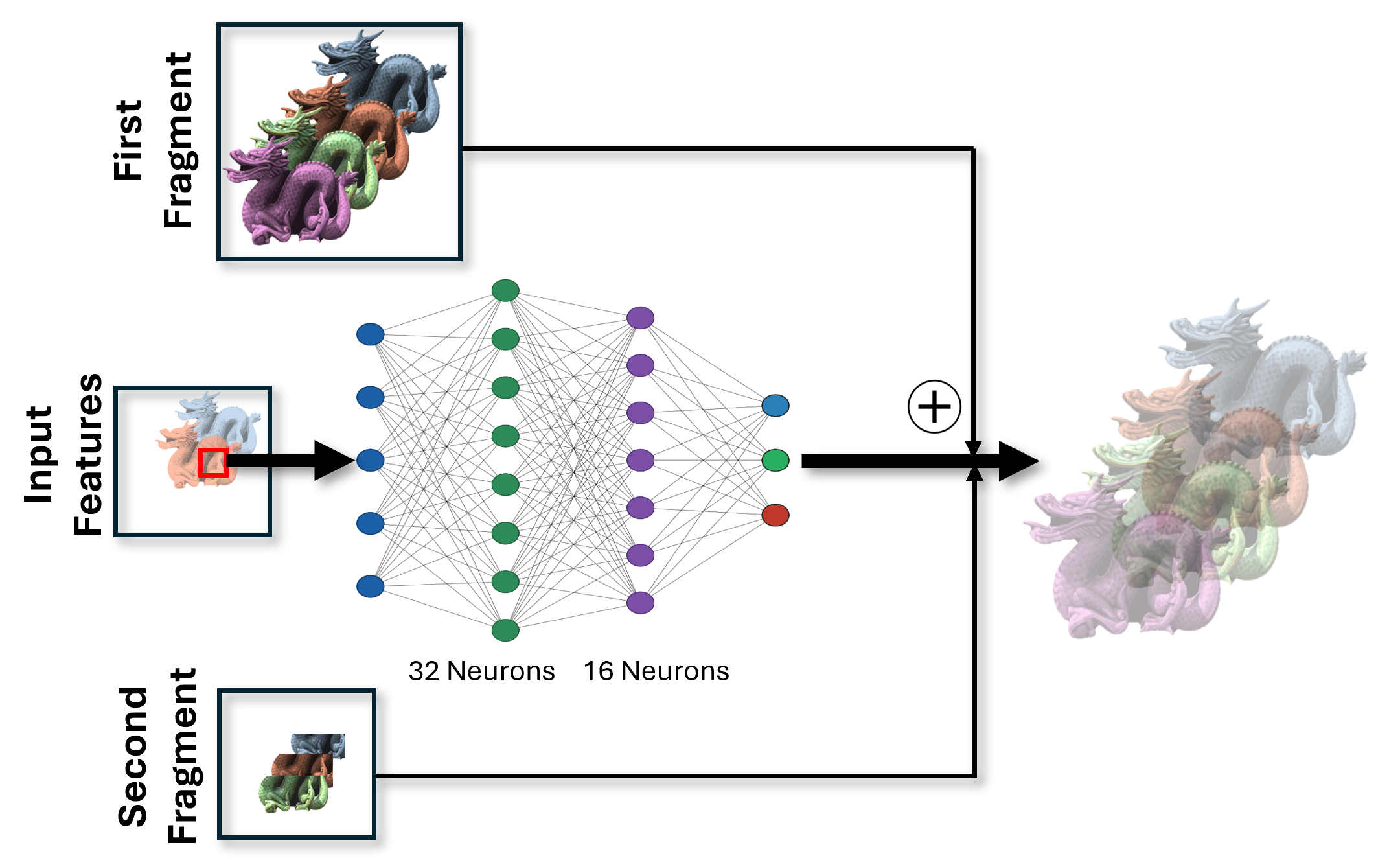}
    \caption{Our neural network predicts the transparency tail color using an adaptive resolution feature pass. The tail color is then blended with the first two fragments, compositing the final OIT color.}
    \label{fig:neuralnetworkfig}
\end{figure}

We build upon the low-end hardware variant of DFAOIT~\cite{tsopouridis2024dfa}, which uses depth peeling to extract the two nearest fragments and an additional geometry pass to accumulate per-pixel statistics without pixel synchronization, making it suitable for mobile and legacy hardware. 


Our neural network is a compact MLP with two hidden layers (32 and 16 neurons) and a 3-neuron ReLU output predicting the tail RGB color. Following DFAOIT~\cite{tsopouridis2024dfa}, the architecture differs only in the output activation (ReLU instead of sigmoid) and predicts only the transparent tail rather than the full transparency color. The network is trained offline on 10 million transparent samples using MSE loss and Adam.

The original DFAOIT \cite{tsopouridis2024dfa} directly predicts the transparency color $C_t$ using 10 input features, related to pixel opacity and color:
\begin{equation}
    C_p = C_t + C_b \prod_{i=1}^{n}(1 - \alpha_i).
\end{equation}
The 10 features are: the exact Porter-Duff blended color of the two front
fragments, the average color $C_{\mathrm{avg}}$ and average opacity $\alpha_{\mathrm{avg}}$ of the remaining $n-2$ fragments, the accumulated premultiplied color $C_{\mathrm{acc}}$, and the fragment count $n$.



We address two limitations of DFAOIT: mixing the exact contribution of the first two depth-peeled fragments with the approximate tail contribution, and the use of unnormalized $C_{\mathrm{acc}}$, which can cause color artifacts. To overcome these issues, the network predicts only the \emph{tail color}, decoupling the exact two-fragment term from the learned approximation:
\begin{equation}
    C_p = C_{\mathrm{exact},2} + T_2 \cdot f(\mathbf{x}) +
          \prod_{i=1}^{n} (1 - a_i) C_b
\end{equation}

\noindent where $C_{\mathrm{exact},2} = \alpha_1 C_1 + (1-\alpha_1)\alpha_2
C_2$ is the exact two-layer composite, $T_2 = (1-\alpha_1)(1-\alpha_2)$ is the transmittance through the two peeled layers, $f(x)$ is the inferred tail color, and $\prod_{i=1}^{n} (1 - a_i)$ is the transmittance of all transparent fragments.

The network utilizes a five-float input vector:
\begin{itemize}
    \item Total fragment count: $N$.
    \item  Tail transmittance: $t_{\mathrm{tail}} = \alpha_{\mathrm{acc}} / T_2$.
    \item Normalized premultiplied RGB color of the tail fragments:
    $\tilde{\mathbf{C}}_{\mathrm{pm}} = C_{\mathrm{acc}} / \alpha_{\mathrm{acc}}$,
    where $C_{\mathrm{acc}} = \sum_{i=3}^{N} \alpha_i \mathbf{C}_i$
    is the premultiplied color accumulated over the tail fragments.
\end{itemize}

The reduction from 10 to 5 features decreases weight tensor memory and
inference cost, improving suitability for constrained hardware (Tab. \ref{tab:passbreakdown}), while tail normalization addresses the color artifacts identified in the original method (See supplemental material).

Geometry passes dominate runtime cost. To reduce this overhead under smooth camera motion, we optionally reuse previous frames through reprojection-based interpolation. Frames alternate between \emph{real renders}, where the full pipeline is executed, and \emph{interpolated frames}, where geometry passes and network inference are skipped. The interpolation interval is adapted at runtime according to the measured frame rate and constrained to a maximum of four frames. When performance drops below the target frame rate, the method reverts to standard rendering without temporal interpolation. Interpolated frames are generated by reprojecting pixels using the nearest available depth and the previous view-projection matrix. Similar to common real-time rendering approaches, this lightweight optimization prioritizes performance and may degrade under rapid motion or highly dynamic transparent geometry.

\section{Experimental Evaluation}


We evaluate STAR-NT against DFAOIT \cite{tsopouridis2024dfa} and WBOIT \cite{McGuire2013} on ten scenes with varying depth complexity $D$ (up to 136 fragments per pixel), color, and opacity, ranging from low to high complexity. WBOIT is used as the baseline OIT method due to its strong real-time performance and plausible quality, using the original authors recommended weight function~\cite{McGuire2013}. Experiments are run at $1920\times1080$ on three representative platforms: Huawei Mate 70 Pro (mobile), NVIDIA GTX 1660 Super (legacy desktop), and NVIDIA RTX 5080 (high-end desktop). Image quality is evaluated using MSE and FLIP~\cite{flip-paper}, where lower values indicate closer agreement with the A-buffer ground truth.

\subsection{Quality}
\label{sec:quality}
\begin{figure*}
    \centering
    \includegraphics[width=0.85\linewidth]{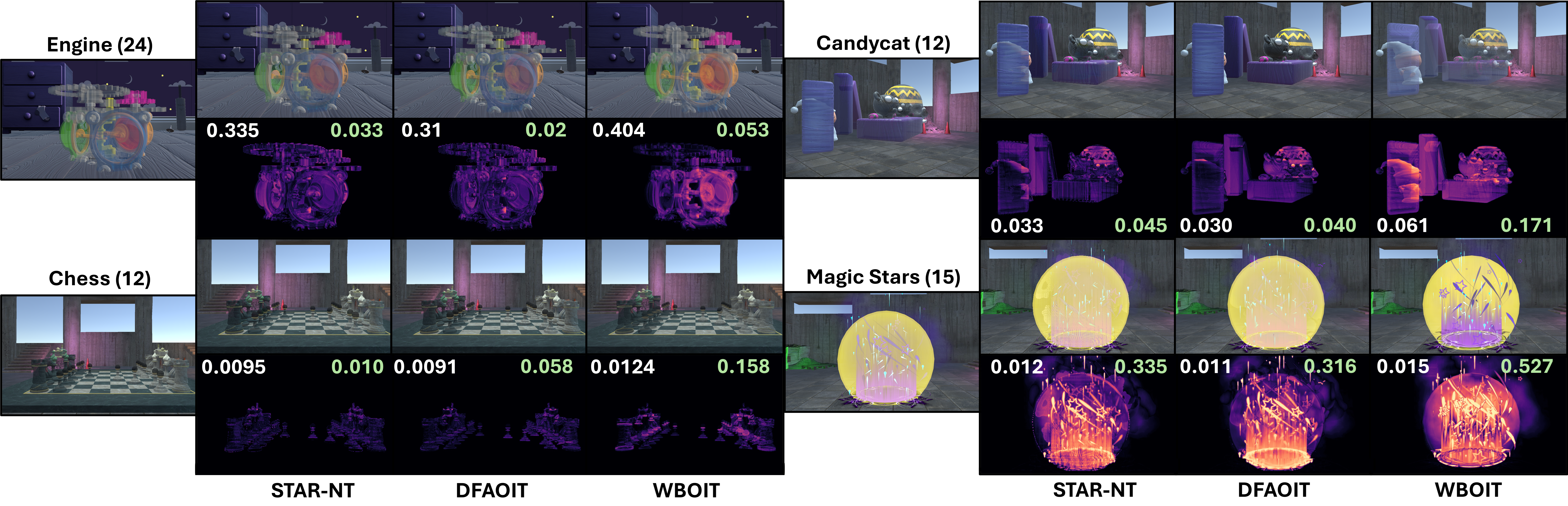}
    \caption{Quality comparison of methods: $FLIP$ mean error (white) and $MSE * 10^2$ (green). The ground truth is shown on the left with the depth complexity (maximum number of transparent layers) in the parenthesis.}
        \label{fig:qualitycomp1}
\end{figure*}

Figure~\ref{fig:qualitycomp1} and the supplementary material report MSE and FLIP errors for all scenes against the A-buffer~\cite{yang2010}, which serves as ground truth. STAR-NT achieves lower error than WBOIT on both metrics, with average MSE and FLIP reductions of 32.2\% and 30.2\%, respectively. DFAOIT achieves slightly lower error than STAR-NT in most scenes, as STAR-NT uses reduced-resolution geometry passes.


Hairball and Powerplant represent challenging high-depth-complexity cases (Supplementary Material). In Hairball, WBOIT achieves the lowest error due to the extremely large fragment count, where STAR-NTs tail prediction operates on more heavily aggregated statistics, while WBOITs weighted blending is better suited for this scene. DFAOIT exhibits visible color artifacts, which are mitigated by our revised inputs. In Powerplant ($D = 96$), STAR-NT achieves the best overall quality, while DFAOIT performs worse than WBOIT, confirming that the improved input representation and decoupled compositing are particularly beneficial in these scenes.

Excluding these two scenes, STAR-NT achieves 40\% and 42\% lower FLIP and MSE, respectively, relative to WBOIT, compared to 50\% and 53\% for DFAOIT. Despite operating with reduced-resolution geometry passes and a smaller feature set, STAR-NT maintains competitive quality while mitigating the color artifacts and instability observed in DFAOIT for high-depth-complexity scenes. Minor artifacts mainly stem from downsampled inputs. Likewise, fast motion and disocclusions are well-known limitations of temporal reconstruction methods. Since the temporal interpolation stage is modular, it can be replaced with more advanced schemes.

We evaluate our method on three platforms and ten scenes with varying depth complexity. STAR-NT achieves consistent frame-time reductions over DFAOIT, with an average speedup of 39\%, with the largest gains on legacy hardware where geometry-pass cost dominates (Tab.~\ref{tab:my-table}). STAR-NT matches the runtime of WBOIT, a single-pass non-neural baseline, while maintaining substantially higher image quality, making neural transparency rendering competitive in performance without sacrificing its quality advantages.

\vspace{-0.2in}
\subsubsection{Improvement Breakdown}

\begin{table*}[t]
\resizebox{\textwidth}{!}{%
\begin{tabular}{|c|c|c|c|c|c|c|c|c|c|c|c|c|}
\hline
\rowcolor[HTML]{EFEFEF}
\textbf{\begin{tabular}[c]{@{}c@{}}Scene \\(P1\%,P2\%,P3\%, D)\end{tabular}}
&
  \textbf{Pass 1} &
  \textbf{Pass 2} &
  \textbf{Pass 2↓} &
  \textbf{Pass 3} &
  \textbf{Pass 3↓} &
  \textbf{Compute} &
  \textbf{\begin{tabular}[c]{@{}c@{}}Inf. \\DFAOIT\end{tabular}} &
  \textbf{\begin{tabular}[c]{@{}c@{}}Inf. \\STAR\end{tabular}} &
  \textbf{\begin{tabular}[c]{@{}c@{}}Total \\DFAOIT\end{tabular}} &
  \textbf{\begin{tabular}[c]{@{}c@{}}Total STAR-NT\\w/o Temporal\end{tabular}} &
  \textbf{\begin{tabular}[c]{@{}c@{}}Total \\STAR\end{tabular}} &
  \textbf{Speedup} \\ \hline
\textit{Smoke (100,50,50, 56)}      & 0.042 & 0.068 & 0.024 & 0.138 & 0.038 & 0.052 & 0.114 & 0.086 & 0.362 & 0.242 & 0.194 & 46.27\% \\ \hline
\textit{Hairball (100,50,50, 136)}   & 0.425 & 0.559 & 0.323 & 1.000 & 0.415 & 0.100 & 0.160 & 0.068 & 2.144 & 1.331 & 0.770 & 64.08\% \\ \hline
\textit{Candycat (100,50,25, 12)}   & 0.048 & 0.040 & 0.025 & 0.089 & 0.020 & 0.071 & 0.102 & 0.078 & 0.279 & 0.242 & 0.210 & 24.73\% \\ \hline
\end{tabular}%
}
\caption{Ablation and per-pass GPU timings (ms) for three scenes at varying adaptive resolution scales (P1\%/P2\%/P3\%). Pass~N↓ denotes reduced-resolution STAR passes. Compute includes tile evaluation and upsampling. Inference reports neural network inference time for the original and redesigned networks. All timings are in ms, $D$ denotes maximum depth complexity.}
\label{tab:passbreakdown}
\end{table*}

Table~\ref{tab:passbreakdown} shows a per-pass timing breakdown for low (Candycat), medium (Smoke), and high (Hairball) complexity scenes on an RTX 5080. The main improvement comes from adaptive resolution scaling in geometry Pass 2 and Pass 3, with the largest gains in Hairball and Smoke due to high geometry cost and overdraw. The redesigned network inputs (Sec.~\ref{neuralnetwork}) provide an additional ~35\% inference speedup by reducing input features from 10 to 5. Temporal interpolation further amortizes full geometry passes across frames.
\vspace{-0.2in}
\subsubsection{Desktop Performance}


Table~\ref{tab:my-table} reports frame times on the RTX~5080. STAR-NT achieves an average 36.8\% reduction over DFAOIT across all scenes, with the largest gains in geometry-heavy workloads. Relative to WBOIT, STAR-NT is only 4.4\% slower on average while providing substantially higher image quality (Sec.~\ref{sec:quality}).


\vspace{-0.2in}
\subsubsection{Mobile Performance} Table~\ref{tab:my-table} reports frame times on the Huawei Mate~70~Pro. STAR-NT reduces frame time over DFAOIT by an average of 28\%, with the largest gains in geometry-heavy scenes such as Tree and Smoke. Powerplant and Hairball show smaller improvements as temporal interpolation is not active due to low frame rates. Relative to WBOIT, STAR-NT is faster in six out of ten scenes, with a 5.5\% average advantage. WBOIT remains faster in low-depth-complexity scenes due to its single-pass design, which is difficult to match with a multi-stage pipeline.

\subsubsection{Legacy Desktop Performance} The GTX~1660~Super results demonstrate an average frame time reduction of 54.0\% relative to DFAOIT. The gains are consistent across all scenes and largest in high depth complexity scenes: Hairball and Powerplant see reductions exceeding 60\% and 64\% respectively. Relative to WBOIT, STAR-NT shows a similar performance, only 0.5\% faster, effectively matching a single-pass approximate method while delivering substantially better image quality. 

\begin{table}[]
\centering
\resizebox{0.7\columnwidth}{!}{%
\begin{tabular}{|l|ccc|ccc|ccc|}
\hline
\rowcolor[HTML]{EFEFEF} 
\multicolumn{1}{|c|}{\cellcolor[HTML]{EFEFEF}\textbf{Scene (D)}} &
  \multicolumn{3}{c|}{\cellcolor[HTML]{EFEFEF}\textbf{DFAOIT}} &
  \multicolumn{3}{c|}{\cellcolor[HTML]{EFEFEF}\textbf{STAR-NT}} &
  \multicolumn{3}{c|}{\cellcolor[HTML]{EFEFEF}\textbf{WBOIT}} \\ \hline
 &
  \multicolumn{1}{l|}{\textit{Mobile}} &
  \multicolumn{1}{l|}{\textit{Legacy}} &
  \multicolumn{1}{l|}{\textit{Desktop}} &
  \multicolumn{1}{l|}{\textit{Mobile}} &
  \multicolumn{1}{l|}{\textit{Legacy}} &
  \multicolumn{1}{l|}{\textit{Desktop}} &
  \multicolumn{1}{l|}{\textit{Mobile}} &
  \multicolumn{1}{l|}{\textit{Legacy}} &
  \multicolumn{1}{l|}{\textit{Desktop}} \\ \hline
Engine (24) &
  \multicolumn{1}{c|}{16.7} &
  \multicolumn{1}{c|}{3.3} &
  0.8 &
  \multicolumn{1}{c|}{\cellcolor[HTML]{DFF2E7}11.1} &
  \multicolumn{1}{c|}{\cellcolor[HTML]{DFF2E7}1.56} &
  \cellcolor[HTML]{DFF2E7}0.54 &
  \multicolumn{1}{c|}{12.5} &
  \multicolumn{1}{c|}{1.6} &
  0.58 \\ \hline
Hair (80) &
  \multicolumn{1}{c|}{11.8} &
  \multicolumn{1}{c|}{2.6} &
  0.78 &
  \multicolumn{1}{c|}{9.1} &
  \multicolumn{1}{c|}{1.44} &
  0.57 &
  \multicolumn{1}{c|}{\cellcolor[HTML]{DFF2E7}8.5} &
  \multicolumn{1}{c|}{\cellcolor[HTML]{DFF2E7}1.2} &
  \cellcolor[HTML]{DFF2E7}0.44 \\ \hline
Magic Stars (15) &
  \multicolumn{1}{c|}{14.1} &
  \multicolumn{1}{c|}{3.8} &
  1.07 &
  \multicolumn{1}{c|}{\cellcolor[HTML]{DFF2E7}11.2} &
  \multicolumn{1}{c|}{\cellcolor[HTML]{DFF2E7}1.7} &
  \cellcolor[HTML]{DFF2E7}0.57 &
  \multicolumn{1}{c|}{11.4} &
  \multicolumn{1}{c|}{1.9} &
  0.61 \\ \hline
Chess (12) &
  \multicolumn{1}{c|}{16.7} &
  \multicolumn{1}{c|}{4.2} &
  0.9 &
  \multicolumn{1}{c|}{\cellcolor[HTML]{DFF2E7}10.5} &
  \multicolumn{1}{c|}{\cellcolor[HTML]{DFF2E7}1.75} &
  0.6 &
  \multicolumn{1}{c|}{13.9} &
  \multicolumn{1}{c|}{1.9} &
  \cellcolor[HTML]{DFF2E7}0.52 \\ \hline
Clothes (13) &
  \multicolumn{1}{c|}{13.9} &
  \multicolumn{1}{c|}{2.4} &
  0.75 &
  \multicolumn{1}{c|}{9.1} &
  \multicolumn{1}{c|}{1.4} &
  0.57 &
  \multicolumn{1}{c|}{\cellcolor[HTML]{DFF2E7}7.7} &
  \multicolumn{1}{c|}{\cellcolor[HTML]{DFF2E7}1.2} &
  \cellcolor[HTML]{DFF2E7}0.5 \\ \hline
Tree (58) &
  \multicolumn{1}{c|}{23.3} &
  \multicolumn{1}{c|}{3.9} &
  0.85 &
  \multicolumn{1}{c|}{\cellcolor[HTML]{DFF2E7}11.8} &
  \multicolumn{1}{c|}{\cellcolor[HTML]{DFF2E7}1.75} &
  0.68 &
  \multicolumn{1}{c|}{17.5} &
  \multicolumn{1}{c|}{1.9} &
  \cellcolor[HTML]{DFF2E7}0.65 \\ \hline
Smoke (56) &
  \multicolumn{1}{c|}{14.9} &
  \multicolumn{1}{c|}{3.0} &
  - &
  \multicolumn{1}{c|}{\cellcolor[HTML]{DFF2E7}8.5} &
  \multicolumn{1}{c|}{1.51} &
  - &
  \multicolumn{1}{c|}{12.5} &
  \multicolumn{1}{c|}{\cellcolor[HTML]{DFF2E7}1.4} &
  - \\ \hline
Candycat (12) &
  \multicolumn{1}{c|}{13.3} &
  \multicolumn{1}{c|}{2.5} &
  - &
  \multicolumn{1}{c|}{10.0} &
  \multicolumn{1}{c|}{1.45} &
  - &
  \multicolumn{1}{c|}{\cellcolor[HTML]{DFF2E7}8.3} &
  \multicolumn{1}{c|}{\cellcolor[HTML]{DFF2E7}1.24} &
  - \\ \hline
Powerplant (98) &
  \multicolumn{1}{c|}{55.6} &
  \multicolumn{1}{c|}{12.3} &
  3.45 &
  \multicolumn{1}{c|}{52.6} &
  \multicolumn{1}{c|}{4.4} &
  \cellcolor[HTML]{DFF2E7}1.1 &
  \multicolumn{1}{c|}{\cellcolor[HTML]{DFF2E7}47.6} &
  \multicolumn{1}{c|}{\cellcolor[HTML]{DFF2E7}4.2} &
  1.25 \\ \hline
Hairball (136) &
  \multicolumn{1}{c|}{58.8} &
  \multicolumn{1}{c|}{11.3} &
  - &
  \multicolumn{1}{c|}{\cellcolor[HTML]{DFF2E7}47.6} &
  \multicolumn{1}{c|}{\cellcolor[HTML]{DFF2E7}2.7} &
  - &
  \multicolumn{1}{c|}{52.6} &
  \multicolumn{1}{c|}{4.7} &
  - \\ \hline
\end{tabular}%
}
\caption{Frame times (ms) on Mobile (Huawei Mate 70 Pro), Legacy (NVIDIA GTX 1660 Super), Desktop (RTX 5080), for scenes of varying maximum depth complexity D.}
\label{tab:my-table}
\end{table}

\vspace{-0.5in}
\subsection{Power and Energy Consumption}
Table~\ref{powertab} reports frame counts, total power consumption, and energy per frame measured on the Huawei Mate~70~Pro across three representative scenes. We used Huawei SmartPerf to measure current draw in milliamps (mA) under sustained rendering, with all methods running at identical scene, display, and battery conditions.


Total current draw or power is not an indicative energy efficiency metric when methods produce different frame counts; energy per frame provides a fair basis for comparison. The energy in $mJ$ for a frame is given by $EN= I \times F_t \times V_d$, where $I$ is the current in $mA$, $F_t$ is the frame duration and $V_d$ is the device voltage. STAR-NT consistently consumes less energy per frame than DFAOIT across all three scenes. Relative to WBOIT, STAR-NT consumes less energy per frame in higher-complexity scenes where neural inference overhead is outweighed by the reduction in geometry pass cost, while drawing slightly more in simpler scenes and delivering substantially better image quality (Section~\ref{sec:quality}).

To jointly assess image quality and energy efficiency, we define a quality-efficiency metric $Q$:
\begin{equation}
    Q = \frac{E_{\mathrm{random}} - E_{\mathrm{method}}}{EN_{\mathrm{frame}}
    \times E_{\mathrm{random}}} \times 100
\end{equation}
\noindent where $E_{\mathrm{method}}$ is the per-scene MSE of each method,
$E_{\mathrm{random}}$ is the MSE of a naive GPU renderer
with no OIT, using the "over" operator on random fragment order, and $EN_{\mathrm{frame}}$ is the energy consumption per frame in $mJ$. $Q$ expresses the percentage of improvement versus the naive-blend error (without sorting) gained per $mJ$ of energy consumed. For example if $Q=1\%$ per $mJ$ for a method this means that each $mJ$ consumed improves the MSE by 1\%.

STAR-NT achieves the highest $Q$ in all three measured scenes (Table~\ref{powertab}), confirming that its performance gains are not offset by quality degradation. WBOIT scores substantially lower as it offers worse quality and in some cases higher energy per frame. DFAOIT underperforms STAR-NT despite the marginally better raw MSE, due to its higher per-frame energy cost. Overall, STAR-NT delivers the best quality improvement per $mJ$ consumed
across all measured scenes.

\begin{table}[]
    \centering

\resizebox{0.4\columnwidth}{!}{%
\begin{tabular}{|l|lrrrr|}
\hline
\rowcolor[HTML]{EFEFEF} 
\multicolumn{1}{|c|}{\cellcolor[HTML]{EFEFEF}\textbf{Scene}} &
  \multicolumn{1}{c|}{\cellcolor[HTML]{EFEFEF}\textbf{Method}} &
  \multicolumn{1}{c|}{\cellcolor[HTML]{EFEFEF}\textbf{FPS}} &
  \multicolumn{1}{p{11mm}|}{\cellcolor[HTML]{EFEFEF}\textbf{Overall power (mW)}} &
  \multicolumn{1}{p{14mm}|}{\cellcolor[HTML]{EFEFEF}\textbf{Energy (mJ) per frame}} &
  \multicolumn{1}{p{15mm}|}{\cellcolor[HTML]{EFEFEF}\textbf{Q \newline \% per mJ}} \\ \hline
\textbf{Smoke}     & \multicolumn{5}{l|}{}                                                                                                  \\ \hline
                   & \multicolumn{1}{l|}{WBOIT} & \multicolumn{1}{r|}{80}  & \multicolumn{1}{r|}{7400} & \multicolumn{1}{r|}{92,50}  & 0,25 \\ \cline{2-6} 
                   & \multicolumn{1}{l|}{DFA}   & \multicolumn{1}{r|}{67}  & \multicolumn{1}{r|}{8140} & \multicolumn{1}{r|}{121,49} & 0,36 \\ \cline{2-6} 
\multirow{-3}{*}{} & \multicolumn{1}{l|}{STAR-NT}  & \multicolumn{1}{r|}{117} & \multicolumn{1}{r|}{8510} & \multicolumn{1}{r|}{72,74}  & 0,55 \\ \hline
\textbf{Candycat}  & \multicolumn{5}{l|}{}                                                                                                  \\ \hline
                   & \multicolumn{1}{l|}{WBOIT} & \multicolumn{1}{r|}{120} & \multicolumn{1}{r|}{6845} & \multicolumn{1}{r|}{57,04}  & 0,93 \\ \cline{2-6} 
                   & \multicolumn{1}{l|}{DFA}   & \multicolumn{1}{r|}{75}  & \multicolumn{1}{r|}{7215} & \multicolumn{1}{r|}{96,20}  & 0,92 \\ \cline{2-6} 
\multirow{-3}{*}{} & \multicolumn{1}{l|}{STAR-NT}  & \multicolumn{1}{r|}{100} & \multicolumn{1}{r|}{7400} & \multicolumn{1}{r|}{74,00}  & 1,18 \\ \hline
\textbf{Engine}    & \multicolumn{5}{l|}{}                                                                                                  \\ \hline
                   & \multicolumn{1}{l|}{WBOIT} & \multicolumn{1}{r|}{80}  & \multicolumn{1}{r|}{6845} & \multicolumn{1}{r|}{85,56}  & 0,68 \\ \cline{2-6} 
                   & \multicolumn{1}{l|}{DFA}   & \multicolumn{1}{r|}{60}  & \multicolumn{1}{r|}{7030} & \multicolumn{1}{r|}{117,17} & 0,64 \\ \cline{2-6} 
\multirow{-3}{*}{} & \multicolumn{1}{l|}{STAR-NT}  & \multicolumn{1}{r|}{90}  & \multicolumn{1}{r|}{7215} & \multicolumn{1}{r|}{80,17}  & 0,93 \\ \hline
\end{tabular}%

}
\caption{Power consumption and quality-efficiency on the Huawei Mate 70 Pro. $Q$ provides the improvement of quality offered by each $mJ$ consumed.}
\label{powertab}
\end{table}
\vspace{-0.5in}
\section{Conclusions}
We present STAR-NT, a dual-domain acceleration framework for neural OIT combining adaptive quadtree-based resolution scaling in geometry passes with depth-based temporal reprojection. This reduces spatial and temporal redundancy, yielding an average 39\% speedup over DFAOIT and up to 64\% on legacy hardware, while requiring no scene-specific preprocessing. The redesigned network inputs improve stability in high-complexity scenes and reduce inference cost. Across platforms, STAR-NT matches the performance of simpler approximate methods while maintaining higher image quality, narrowing the gap between neural transparency and real-time deployment. Future work includes learned frame synthesis, spatially selective inference, and hardware-aware acceleration.

 \bibliographystyle{splncs04}
 \bibliography{paper48}

\end{document}